\documentclass[12pt]{article}
\usepackage[cp1251]{inputenc}
\usepackage[english]{babel}
\renewcommand{\thesection}{\Roman{section}}
\def\theequation{\arabic{section}.\arabic{equation}}
\begin{document}

\title{The variational principle and effective action for a spherical dust
shell}
\date{\today}
\author{Valentin~D.~Gladush \\
{\slshape Department of Physics,  Dnepropetrovsk National University,}
\\ {\slshape per. Nauchniy 13, Dnepropetrovsk 49050, Ukraine}
\\ E-mail: gladush@ff.dsu.dp.ua} 

\maketitle

\begin{abstract}
The variational principle for a spherical configuration consisting of
a thin spherical dust shell in gravitational field is constructed.
The principle is consis\-tent with the boundary-value problem of the
corresponding Euler-Lagrange equations, and leads to ``natural
boundary conditions''. These conditions and the field equations
following from the variational principle are used for perform\-ing of
the reduction of this system. The equations of motion for the shell
follow from the obtained reduced action. The transformation of the
variational formula for the reduced action leads to two natural
variants of the effective action. One of them describes the shell
from a stationary interior observer's point of view, another from the
exterior one. The conditions of isometry of the exterior and interior
faces of the shell lead to the momentum and Hamiltonian constraints.
\end{abstract}

\section{Introduction}
\setcounter{equation}{0} A spherically-symmetric dust shell is among
the simplest popular models of collapsing gravitating configurations.
The equations of motion for these objects were obtained in Refs.
\cite{israel1} and \cite{kuchar1}. The construction of a variational
principle for such systems was discussed in Refs.
\cite{visser}--\cite{berezin3}. There are a number of problems here,
basic of which is a choice of the evolution parameter (internal,
external, proper). The choice of time coordinate, in turn, affects
the choice of a particular quantization scheme, leading, in general,
to quantum theories which are not unitarily equivalent.

In most of these papers the variational principle for shells is
usually constructed in a comoving frame of reference, or in one of
variants of freely falling frames of reference. However, using of
such frames of reference frequently leads to effects unrelated to the
object under consideration. In the approach related to proper time of
the shell reduction of the system leads to complicated Lagrangians
and Hamiltonians which creates difficulties on quantization. In
particular it leads to theories with higher derivatives or to finite
difference equations.

The essential physics involves a picture of a gravitational collapse
from the point of view of an infinitely remote stationary observer.
In quantum theory this point of view enables us to treat bound states
in terms of asymptotic quantities and to build the relevant
scattering theory correctly. On the other hand, to treat primordial
black holes in the theory of self-gravitating shells it is convenient
to take the viewpoint of a central stationary observer.

In our opinion, the choice of exterior or interior stationary
observers is most natural and corresponds to real physics. The
natural Hamiltonian formulation of a self-gravitating shell was
considered in Refs. \cite{hajicek5}, \cite{dolgov}. However, this
formulation was not obtained by a variational procedure from some
initial action containing the standard Einstein-Hilbert term.

The general Lagrange approach to the theory of dust shells in General
Relativity was developed in Ref. \cite{gl3}. In present paper the
natural Lagrange and Hamiltonian formulation of the spherical
self-gravitating dust shell is constructed, which has some specific
features in comparison with general approach. The system under
consideration is regarded as a compound spherical configuration
consisting of two vacuum spherical regions $D_{-}$ and $D_{+}$ with
spherical boundary surface  $\Sigma$ formed by the shell. The initial
action is taken as the sum of actions of the York type $I_{Y}=
I_{EH}+ I_{\partial D}$ \cite{york1} for each regions and the action
for the dust matter on the singular hypersurface $\Sigma$. Here
$I_{EH}$ is the the Einstein--Hilbert action, and $I_{\partial D}$ is
the boundary term. The constructed variational principle is
compatible with boundary-value problems of the corresponding
Euler--Lagrange equations for each region of the configuration, and,
when we vary with respect to metric, leads to the ``natural boundary
conditions'' on the shell. The obtained conditions together with
gravitation field equations, are used for performing the reduction of
the system and eliminating of the gravitational degrees of freedom.
The equation of motion for the shell is obtained from the reduced
action by considering normal variations of the shell.

Transforming of the variational formula and applying of the surface
equations leads to two variants of effective action. One of them describes
the shell from an interior stationary observer's point of view, and the
other from the exterior one. Going over to the Hamiltonian description and
using the isometry conditions of the exterior and interior faces of the
shell generates momentum and Hamiltonian constraints.

Here $c$ is the velocity of light, $k $ is the gravitational constant,
$\chi =8\pi k/c^2$. The metric tensor $g_{\mu\nu}\, (\mu,\nu=0,1,2,3)$ has
signature (+~-~-~-).

\section{Total action for the configuration, bulk\\
and surface equations}
\setcounter{equation}{0}

Consider a set of the regions $D=D_{-} \cup \Sigma \cup D_{+}\subset
V^{(4)}$ in spherically symmetric space-time $V^{(4)}$. Here $D_{-}$ and
$D_{+}$ are the interior and exterior regions, respectively, which are
separated by the spherically-symmetric infinitely thin dust shell $\Sigma$
with the surface dust density $\sigma$. Choose in $D_{\pm}$ the general
angle coordinates $x^i: \{x^2=\theta,~x^3=\alpha\}\ (i,k=2,3)$ and
individual space-time coordinates $x^{a}_{\pm} \ (a,b=0,1)$ for $D_{\pm}$,
respectively. Then the gravitational fields in the regions $D_{\pm}$ are
described by the metrics
\begin{equation}\label{2.1}
      ~^{(4)}ds^{2}_{\pm}=~^{(2)}ds^{2}_{\pm}-r^2 d\sigma^2\,,
\end{equation}
\begin{equation}\label{2.2a}
  ~^{(2)}ds^{2}_{\pm}=  \gamma^{\pm}_{ab}dx^{a}_{\pm}dx^{b}_{\pm}\,, \quad
  d\sigma^2=h_{ij}dx^{i}dx^{j}=d\theta^2+\sin^2 \theta d\alpha^2\,,
\end{equation}
where the two-dimensional metrics $\gamma^{\pm}_{ab}$ and the function $r$
depend on the coordinates $x^{a}_{\pm}$.

Einstein's equations and  the curvature scalar for each region $D_{\pm}$
can be represented in the form
\begin{eqnarray}
    ~^{(4)}G_{ab} & = & -\frac{2}{r}\,\nabla_{a}\nabla_{b}r
    +\frac{1}{r^2}\left(2r\Delta ~r -(\nabla r)^2 +1
    \right)\gamma_{ab}=0 \, ,     \label{gab}\\
    ~^{(5)}G_{aj} & \equiv & 0\, ,    \label{gai}\\
    ~^{(4)}G_{ij} & = &\frac{r}{2}\left(r~^{(2)}R
    -2\Delta r \right)h_{ij}=0\,,   \label{gik}
\end{eqnarray}
\begin{equation}\label{ric}
      ~^{(4)}R=~^{(2)}R - \frac{4}{r}~\Delta r -
      \frac{2}{r^2}(\nabla r)^2 - \frac{2}{r^2}\, ,
\end{equation}
where $~\Delta r =\nabla^{a}\nabla_{a}r = r^{\,,\,a}_{\,;\,a}, \
(\nabla r)^2 = \gamma^{ab}\nabla_{a} r \nabla_{b}
r=r^{\,,\,a}r_{,\,a}$, $\nabla_{a}\equiv _{\,;\,a}$ is  the covariant
derivative with respect to $x^a$ in metric $\gamma_{ab}$, and
$^{(2)}R$ is the curvature scalar of two-dimensional space with
metric $\gamma_{ab}$, $r_{,\,a}\equiv {\partial}r/{\partial x^{a}}$.
Here, for simplicity, we temporarily omit the signs ``$_{\pm}$''.

Now we introduce a general coordinate map $x^{a}\in D$, and metrics
$\gamma^{\pm}_{ab}$ such that ${\gamma^{-}_{ab}}_{|_{\Sigma}} =
{\gamma^{+}_{ab}}_{|_{\Sigma}}= \gamma_{ab}$. Then
${~^{(2)}ds_{+}}_{|_{\Sigma}}={~^{(2)}ds_{-}}_{|_{\Sigma}}\equiv ~^{(2)}
ds $, and the world line $\gamma$ of the shell in this map are given by
equation $x^{a}=x^{a}(s)$. Let
\begin{eqnarray}\label{bas}
   \left\{ \vec{u}= u^{a}\partial_{a}\,,\
  \vec{n}= n^{a}\partial_{a} \right\}, \quad
  \left\{\ \omega = u_{a}dx^{a}\,, \  \eta = n_{a}dx^{a}\ \right\}
\end{eqnarray}
be the general orthonormal vector and covector bases in the regions
$D_{\pm}$. Here $\partial_{a} ={\partial}/{\partial x^{a}}$ is the
partial derivative with respect to $x^a$. The components of vectors
$\{\vec{u}, \vec{n}\}$ and covectors $\{\omega, \eta \}$ satisfy the
conditions ${u_{a}u^{a}}_{|\pm} =-{n_{a}n^{a}}_{|\pm} =1$ and
${u_{a}n^{a}}_{|\pm} = 0$. Hence, accurate to a general factor $
\epsilon = \pm 1 $, we obtain
\begin{eqnarray}
  n_{0}=\sqrt{-\gamma}\ u^{1}, \qquad n_{1}=- \sqrt{-\gamma}\
u^{0}\,,\label{n_u} \\
  u_{0}=\sqrt{-\gamma}\ n^{1},
  \qquad u_{1}=- \sqrt{-\gamma}\ n^{0}\,.\label{n_ua}
\end{eqnarray}
where  $\gamma = \det |\gamma_{ab}| $. With respect to the bases
$\{\vec{u}, \vec{n}\}$ and $\{\omega, \eta \}$ we have
\begin{equation}\label{met1}
  \gamma_{ab}=u_a u_b - n_a n_b\,, \quad
  \gamma^{ab}= u^a u^b - n^a n^b\,, \quad
  \delta^{a}_{b}= u^a u_b - n^a n_b\,.
\end{equation}

Further, we shall suppose that the vector field $ \vec {u} $ in points
$p\in\Sigma $ is tangential to a world line of a shell $ \gamma $ so, that
$ {u^{a}_{\pm}}|_{\Sigma} = {dx^{a}}/{^{(2)}ds} $. The vector field $ \vec
{n} $ in points $p\in\Sigma $ is normal to $ \Sigma $ and is directed from
$D_- $ in $D_ + $. Inside regions $D _ {\pm} $ the field of an orthonormal
dyad $\{u^a,\ n^a\}$ is arbitrary.

Define the one-forms $d \Sigma_{a}$  as
\begin{equation}\label{dxdS}
   dx^{a} \wedge d\Sigma_{b} = \delta^{a}_{b}\, dx^0 \wedge dx^1
  = \delta^{a}_{b}\,d^2 x\,,
\end{equation}
where the symbol ``$\wedge$''  denotes the exterior product. It is also
useful to define the one-forms
\begin{eqnarray}\label{vol1}
   d\Sigma_{u}= u^{a} d \Sigma_{a}\, , \quad
   d\Sigma_{n} = - n^{a} d \Sigma_{a}\,,
\end{eqnarray}
which are dual to the one-forms $\omega\,, \eta $, so that
\begin{eqnarray}
    & \sqrt{-\gamma}\,\omega\wedge d\Sigma_{u}
   = \sqrt{-\gamma}\,\eta\wedge d\Sigma_{n}  =\sqrt{-\gamma}\,d^2 x =
   \omega\wedge\eta\,,\label{vol2} \\
 & \omega = -\sqrt{-\gamma}\,d\Sigma_{n}\,, \qquad
 \eta = \sqrt{-\gamma}\,d\Sigma_{u}\,.\label{vol3}
\end{eqnarray}
In addition we have $ \ g = \det |g_{\mu\nu}|=\gamma r^4 \sin^2\theta $.

Now we introduce the tensors of extrinsic curvature
\begin{equation}\label{extcurR}
  K_{\mu\nu} = - n_{\mu\,;\,\rho}(n^{\rho}n_{\nu}+ \delta^{\rho}_{\nu})\,,
  \qquad K=g^{\mu\nu} K_{\mu\nu}= -n^{\mu}_{;\,\mu}\,,
\end{equation}
\begin{equation}\label{extcurT}
  D_{\mu\nu} =  u_{\mu\,;\,\rho}(u^{\rho}u_{\nu} - \delta^{\rho}_{\nu})\,,
  \qquad D=g^{\mu\nu} D_{\mu\nu}= -u^{\mu}_{;\,\mu}\,
\end{equation}
of local subspaces $\Sigma_n$ and $\Sigma_u$, which are orthogonal to
the vectors $n^{\mu}=\{n^{a}, 0 , 0\}$ and $u^{\mu}=\{u^{a}, 0 ,
0\}$, respectively. Here ``$_{;\,\rho}$'' is  the covariant
derivative with respect to metric $g_{\mu\nu}$, $K$ and $D$ are the
trace of the tensors $K_{\mu\nu}$ and $D_{\mu\nu}$, respectively. On
the shell the tensor $K_{\mu\nu}$ is the tensor of extrinsic
curvature hypersurface $\Sigma$.

From the definitions (\ref{extcurR}) and (\ref{extcurT}) we can obtain
\begin{equation}\label{extcurR1}
  K_{i\,k}= r (\vec{n} r) h_{ik}\,, \quad K_{ai}= K_{ia} =0\,, \quad
  K_{ab}=K_{u u}\, u_{a}u_{b}\,,
\end{equation}
\begin{equation}\label{extcurR2}
 K_{u u}=K_{a b}u^{a}u^{b}=-n^{a}_{\,;\,a}\,,
  \quad K = -\frac{1}{r^2}\left(r^2n^{a}\right)_{\,;\,a}
  = K_{u u} -\frac{2(\vec{n} r)}{r} \,,
\end{equation}
\begin{equation}\label{extcurT1}
  D_{i\,k}= r (\vec{u} r) h_{ik}\,, \quad D_{ai}= D_{ia} =0\,, \quad
  D_{ab}= D_{n n}\, n_{a}n_{b}\,,
\end{equation}
\begin{equation}\label{extcurT2}
   D_{n n}=D_{a b}n^{a}n^{b}= u^{a}_{\,;\,a}\,,
  \quad D = -\frac{1}{r^2}\left(r^2 u^{a}\right)_{\,;\,a}
  = - D_{n n} -\frac{2(\vec{u} r)}{r}\,,
\end{equation}
where $\vec{n}r= n^a r_{,\,a}$, $\vec{u}r= u^a r_{,\,a}$.

We take the total action for the spherically symmetric compound
configuration under consideration  in the form
\begin{equation}\label{itot1}
     I_{tot} = I_{EH} + I_{m}+I_{\Sigma} + I_{\partial D} + I_{0}\,,
\end{equation}
where
\begin{equation}\label{ieh}
I_{EH} = - \frac{c}{2\chi} \int\limits_{D_{-\,}\cup D_{+}}
\sqrt{-g}~^{(4)} R~d^2 x \wedge d\theta\wedge d\alpha
\end{equation}
is the sum of Einstein-Hilbert actions for the regions $D_{\pm}$.

The dust on the singular shell $\Sigma$ is described by the action
\begin{equation}\label{idust}
 I_{m} = c\int\limits_{\Sigma}\sigma \sqrt{-g}\
     d\Sigma_{n}\wedge d\theta\wedge d\alpha\,.
\end{equation}
The third term in the right-hand side (\ref{itot1}) is the matching term
\begin{equation}\label{iSigma}
     I_{\Sigma}
  = - \frac{c}{\chi}\int\limits_{\Sigma}
  \sqrt{-g}\,\,[K]\,d\Sigma_{n}\wedge d\theta\wedge d\alpha\,,
\end{equation}
where the symbol $[A]=A|_{+}-A|_{-}$ denotes the jump of the quantity $A$
on the shell $ \Sigma$. The signs ``$|_{\pm}$'' indicate that the marked
quantities are calculated as the limit values when we approach to $\Sigma$
from inside and outside, respectively.

The fourth term in the right-hand side (\ref{itot1})
\begin{equation}\label{4.3}
      I_{\partial D} = \frac{c}{\chi} \oint\limits_{\partial D}
      \sqrt{-g}\,(D u^a -K n^a) d \Sigma_{a}\wedge d\theta\wedge d\alpha
\end{equation}
contains the surface terms similar to Gibbons-Hawking surface term,
which are introduced to fix the metric on the boundary $\partial D$
of the region $ D$. Note, that the boundary $\partial D$  consists of
the pieces of timelike as well as spacelike hypersurfaces. The last
term $I_{0}$ in (\ref{itot1}) contains the boundary terms, necessary
for normalization of the action. It is needed when exterior boundary
$\partial D_+$ of the region $D_+$ is situated on the timelike
infinitely remote hypersurfaces.

Thus the total action $I_{tot}$ is the functional of the metrics
$\{\gamma^{+}_{ab}, \gamma^{-}_{ab}\}$, of the shell radius $r$ and of the
hypersurface $\Sigma$: $ I_{tot}\equiv I_{tot}[\gamma^{+}_{ab},
\gamma^{-}_{ab}, r; \Sigma]$.

The first and the fourth terms in (\ref{itot1}) form the action of the
York type $I_{Y}= I_{EH}+ I_{\partial D}$ \cite{york1}. It is used in
variational problems with the fixed metric on the boundary $\partial D$ of
the configuration $ D$. This action can also be used in variational
problems with the general relativistic version of the ``natural boundary
conditions'' for ``free edge'' \cite{hayward1}, when the metric on the
boundary is arbitrary and the corresponding momenta vanishes. Together
with $I_{0}$ it forms the York--Gibbons--Hawking action $ I_{YGH}= I_{Y}
+I_{0} $ for a free gravitational field.

In our case of the compound configuration we also fix the metric on the
boundary $\partial D$. However, in addition, we have the boundary surface
$\Sigma$ inside the system, with the singular distribution of matter on
it. We can treat this configuration as two vacuum regions $D_{\pm} $ with
common ``loaded edge'' (or ``massive edge'') $\Sigma$. The sum of the
actions of type $I_{Y}$ for these regions, and of the action for matter
$I_{m} $, and normalizing term $I_{0} $ do leads to the action $I_{tot}$.

If there is no dust, $\sigma =0$, the common boundary is not ``loaded''.
Then, the requirement $\delta I_{tot}=0 $, at arbitrary, everywhere
continuous variations of the metric, gives generalization of the above
``natural boundary conditions'' for free hypersurface $ \Sigma $. They
coincide with the continuity conditions for the extrinsic curvature on $
\Sigma $, i.e. with the standard matching conditions. If the matched edges
``are loaded'' by  some surface matter distribution, then we obtain the
surface equations or the boundary conditions for $D_{\pm}$. They are the
analog of the generalized ``natural boundary condi\-tions'' for ``loaded
edges''. The initial action $I_{tot}$ was so chosen, that the surface
equations on $\Sigma$ which follow from requirement $\delta I_{tot}=0$
coincide with the matching conditions on singular hypersurfaces
\cite{israel1}. In that case the variational principle for the action $
I_{tot}$ will be compatible with the boundary-value problem of the
corresponding Euler--Lagrange equations \cite{kurant}, \cite{ponomarev}.

After integrating with respect to angles and taking into account the
relations (\ref{ric}) and (\ref{vol3}), the actions (\ref{ieh}) and
(\ref{idust}) can be written in the form
\begin{equation}
       I_{EH} = - \frac{c^3}{4k} \int\limits_{D^{(2)}_{-}\cup ~D^{(2)}_{+}}
 \sqrt{-\gamma}\left(^{(2)}R\,r^2 - 4r\Delta r -
   2(\nabla r)^2 -2\right)d^2 x\,,\label{ieh2}
\end{equation}
\begin{equation}
       I_{m} = mc\int\limits_{\Sigma^{(1)}}\sqrt{-\gamma}\, d\Sigma_{n}
       = - mc\int\limits_{\gamma}\omega \,,\label{idus2}
\end{equation}
where $m=4\pi\sigma r^2=\mbox{const}$ is the shell mass.

The matching (\ref{iSigma}) and  the boundary surface (\ref{4.3}) terms
can be written as
\begin{eqnarray}
     I_{\Sigma} = \frac{c^3}{2k}\int\limits_{\gamma}
      r^2\,[K]\, \omega
      = \frac{c^3}{2k}\int\limits_{\gamma}
      r\left[r  K_{u u}
  - 2(\vec{n}r) \right]\, \omega\,,\label{iSigma1}\\
    I_{\partial D} = \frac{c^3}{2k}\oint\limits_{\partial D}
     r^2\sqrt{-\gamma}\,(D u^a -K n^a) d \Sigma_{a} \,.\label{iSigmaTR}
\end{eqnarray}

In order to simplify the total action $I_{tot1}$ we reduce the action
(\ref{ieh2}) to the form including only the first-order derivatives. To
this end, we use the fact that, in the two-dimensional space, a curvature
scalar can be reduced  (locally!) to the divergence of a vector (see
Appendix A)
\begin{equation}\label{div}
~^{(2)}R = 2V^{a}_{\,;\,a}\,,
\end{equation}
\begin{equation}\label{V}
       V^{a} = n^b_{;\,b}n^a - u^b_{;\,b}u^a
       = -K_{u u} n^a - D_{n n}u^a\,.
\end{equation}
Then, using the formulae
\begin{equation}\label{div0}
  \sqrt{-\gamma}\,r^2~^{(2)}R
  = 2\sqrt{-\gamma}\,r^{2}V^{a}_{\,;\,a}
  = 2(\sqrt{-\gamma}\,r^{2}V^{a})_{\,,\,a}-4r r_{\,,\,a}V^{a}\,,
\end{equation}
\begin{equation}\label{div1}
     r\sqrt{-\gamma} \Delta r = (\sqrt{-\gamma}\,r r^{\,,\,a})_{,a}
    - \sqrt{-\gamma}\,(\nabla r)^2\,,
\end{equation}
the Einstein-Hilbert actions (\ref{ieh2}) can be rewritten as
\begin{equation}\label{iehg}
  I_{EH}=I_{g} - I_{\partial}\,,
\end{equation}
where
\begin{equation}\label{ig}
      I_{g}= \int\limits_{~D^{(2)}_{-}\cup ~D^{(2)}_{+}}L_{g} d^2 x\,.
\end{equation}
is the gravitational action for the gravitational field with the
Lagrangian, which includes only the first order derivatives
\begin{equation}\label{Lg}
  L_{g}  = \frac{c^3}{2k} \sqrt{-\gamma}\,
   (2r r_{,\,a}V^{a} - r_{,\,a}r^{,\,a} + 1 ) \,.
\end{equation}
Here $r^{,\,a}=\gamma^{ab}r_{,\,a}\,$,  $\ r_{,\,a}=\partial
r/\partial x^a = (\vec{u}{r}) u_{a} -(\vec{n}{r})n_{a}\,$, $\
r_{,\,a}r^{,\,a} = (\vec{u}r)^{2} - (\vec{n}r)^{2}\,$.

The second term in (\ref{iehg}) is the sum of two surface terms
\begin{equation}\label{iD}
    I_{\partial}
    = \frac{c^3}{2k}\, \oint\limits_{\partial D^{(2)}_{-}}
    r\sqrt{-\gamma}\,\, W^a d\Sigma_{a}
  + \frac{c^3}{2k}\, \oint\limits_{\partial D^{(2)}_{+}}
  r\sqrt{-\gamma}\,\, W^a d\Sigma_{a}\,,
\end{equation}
where
\begin{equation}\label{W}
  W^a = r V^a - 2r^{\,,\,a}\,.
\end{equation}
The term (\ref{iD}) includes the integration over total boundaries
$\partial D_{+}$ and $\partial D_{-}$ of the regions $D_{-}$ and $D_{+}$.
Further, take into account (\ref{V}), (\ref{extcurR2}) and
(\ref{extcurT2}), we find
\begin{equation}\label{W1}
  W^a = \left(r D_{n n} -2(\vec{u}r)\right) u^{a}
  -\left(r K_{u u} -2(\vec{n})r\right)n^{a}
  = r\left( D u^{a}- K n^{a}\right)\,.
\end{equation}
Now the term (\ref{iD}) can be rewritten as the sum of two addends
\begin{equation}\label{iD1}
    I_{\partial} = \tilde{I}_{\Sigma} + \tilde{I}_{\partial D}\,.
\end{equation}
The addend $\tilde{I}_{\partial D}$ includes the integration only over
that part of boundaries $\partial D_{+}$ and  $\partial D_{-}$ of region
$D_{-}$  and $D_{+}$  which coincides with the boundary $\partial D $ of
configuration $D=D_{-} \cup \Sigma \cup D_{+}$. In the addend
$\tilde{I}_{\Sigma}$ we integrate over the remaining parts of the
boundaries $\partial D_{\pm}$, which means the integration over exterior
and interior sides of common boundary $\Sigma$ of the regions $D_{+}$ and
$D_{-} $\ , i. e. over the exterior and interior faces of the dust shell.
Taking into account (\ref{iD}), (\ref{W1}), (\ref{iSigma1}) and
\ref{iSigmaTR}) it is easy to see, that $\tilde{I}_{\partial
D}=I_{\partial D}$ and $\tilde{I}_{\Sigma} = I_{\Sigma}$. After
substitution  (\ref{iehg}) and (\ref{iD1}) into (\ref{itot1}), the surface
terms are reduced and complete action acquires the ordinary and natural
form.
\begin{equation}\label{itot2}
     I_{tot} = I_{g}  + I_{m}+ I_{0}\, ,
\end{equation}
where the action $I_{g} $ contains the Lagrangian (\ref{Lg}) with
first-order derivatives only.

The forms (\ref{itot1}) and (\ref{itot2}) of the action $I_{tot}$ are
equivalent. Applying the action (\ref{itot1}), we can evaluate the value
of $I_{tot}$ on the extremals, whereas we use formula (\ref{itot2}) for
finding the extremals.

Now find the variation $\delta I_{tot}$ generated by varying $r$ and
$\gamma^{ab}$. Using relations (\ref{met1}) and
\begin{equation}\label{var-g}
  \delta\sqrt{-\gamma}=-\frac{1}{2}\sqrt{-\gamma}\gamma_{ab}\delta\gamma^{ab}
  =\sqrt{-\gamma}(n_a \delta n^a - u_a \delta u^a)\,,
\end{equation}
it is convenient to express the variations of the metric
$\delta\gamma^{ab}$ through the variations of the vectors $u^a $ and
$n^a $ in the final formulas.

In order to calculate the variation $\delta I_{g}$, we use formulae
\begin{eqnarray}\label{var-r}
      \delta(r_{\,,\,a}r^{\,,a})=2(\dot{r}r_{\,,\,a}\delta u^{a}-
      r'r_{\,,\,a}\delta n^{a}-  r^{\,,\,a}_{\,;\,a}\delta r )
      +2(r^{\,,\,a}\delta r)_{\,;\,a}
\end{eqnarray}
\begin{eqnarray}\label{var-V}
  & \delta(\sqrt{-\gamma}\,rr_{\,,a}V^{a})=
  -\sqrt{-\gamma}\,r V^{a}_{\ ;\,a}\delta r \nonumber\\
  & + \sqrt{-\gamma}(rr_{,a}n^{b}_{;b}+(r\dot{r})^{\cdot}n_{a}
   -(r r')^{\cdot}u_{a})\delta n^a \nonumber\\
  & - \sqrt{-\gamma}(rr_{,a}u^{b}_{;b}+(r\dot{r})'n_{a}
   -(r r')'u_{a})\delta u^a \\
    & + \{r\sqrt{-\gamma}\,V^a \delta r
  + r r'\delta(\sqrt{-\gamma}\,n^a)
  - r\dot{r}\delta(\sqrt{-\gamma}\,u^a)\}_{,\,a}\,, \nonumber
\end{eqnarray}
where $r_{,a}\equiv\partial r/\partial x^a = \dot{r}u_{a}-r'n_{a}$
and $\dot{r}=r_{,a}u^a=\vec{u}\,r\,, \ r'=r_{,a}n^a=\vec{n}\,r $.

Further we assume the boundary of configuration $\partial D$, and
also the metric on it and the normal to be fixed. Therefore $
\delta d\Sigma_{a} |_{\partial D} =0,\ \delta r|_{\partial D}=0,\
\delta u^{a}|_{\partial D}=\delta n^{a}|_{\partial D}=0 $. In
addition, the hypersurface $\Sigma $ is fixed and the metric and
its variations are continuous on $\Sigma$. Hence $~[\gamma^{ab}]
=[~\delta \gamma^{ab}] = 0 $, $ [n^{a}] =[\delta n^{a}] = [u^{a}]
= [\delta u^{a}] =0 $.

According to (\ref{idus2}), for the variation $\delta I_{m}$ we have
\begin{equation}\label{dIm}
  \delta I_{m}= -m c\,\delta\int\limits_{\gamma}\omega =- m c
\int\limits_{\gamma}\delta\omega_{\gamma}\,.
\end{equation}
The sign ``$~_{|\gamma}$'' denotes the restriction of the one-form
$\omega$ on the shell world line $\gamma $: $$ \omega_{\gamma}=
(u_{a}dx^{a})_{\gamma}=(u_{a}dx^{a}/^{(2)}ds)\,^{(2)}ds =
u_{a}u^{a}\,^{(2)}ds =~^{(2)}ds\,, $$ such that
\begin{equation}\label{var-ds}
  \delta\omega_{|\gamma}=\delta ~^{(2)}ds
  =\frac{1}{2}u^{a}u^{b}~^{(2)}ds~\delta \gamma_{ab}
  =- \frac{1}{2}u_{a}u_{b} \omega_{|\gamma}~\delta \gamma^{ab}
  =- u_{a} \omega_{|\gamma}\delta u^{a}\, .
\end{equation}
The requirement of stationarity $\delta I_{tot}=0$ with respect to
arbitrary variations $ \delta u^a $, $\delta n^a$ satisfying the
above-mentioned conditions leads to
\begin{eqnarray}
   \dot{r}' - r'u^{b}_{;b}=0\,,\quad
   \acute{r}^{\cdot} - \dot{r}n^{b}_{;\,b}=0\,, \label{eq-g1}\\
   2r\ddot{r}- 2r r'n^{b}_{;\,b}+ \dot{r}^2-r'^2+1=0\,, \label{eq-g2}\\
    2r{\acute{r}}'- 2r\dot{r}u^{b}_{;\,b}- \dot{r}^2+r'^2-1=0 \label{eq-g3}\,.
\end{eqnarray}
In deriving formulas (\ref{var-V}) - (\ref{eq-g3}) we used
equations
\begin{eqnarray}\label{2dim}
 & u^{a}_{\,;\,b}u^{b} = -n^{a}n_{c}u^{c}_{\,;\,b}u^{b}
  = n^{a}n^{b}_{;\,b}\,, \quad
  u^{a}_{\,;\,b}n^{b} = n^{a}u^{b}_{;\,b}\,, \nonumber \\
 & n^{a}_{\,;\,b}n^{b} = u^{a}u_{c}n^{c}_{\,;\,b}n^{b}
  = u^{a}u^{b}_{;\,b}\,, \quad
   n^{a}_{\,;\,b}u^{b} = u^{a}n^{b}_{;\,b}\,,
\end{eqnarray}
which are specific to the two-dimensional case. It can easily be shown
that the equations (\ref{eq-g1}) - (\ref{eq-g3}) are equivalent to the
equations (\ref{gab}) written in the basis $\{ u^a , n^a\}$.

The variations of $I_{tot}$ with respect to $r$ lead to equation
\begin{equation}\label{eg-r}
  r V^{\,a}_{\,;\,a}-\Delta r=0\,,
\end{equation}
which, in view of (\ref{div}), is equivalent to the rest of the
Einstein equations (\ref{gik}). Besides equations
(\ref{eq-g1})-(\ref{eg-r}) we also obtain the the surface equation
for jumps
\begin{eqnarray}
  &[\vec{n}\,r] - r[n_{a}V^{a}]=0\,, \label{eq-s1}\\
  & c^2 r[\vec{n}\,r] + k m =0\,.\label{eq-s2}
\end{eqnarray}

Note that by virtue of (\ref{V}) and (\ref{2dim}) there exist formulae
\begin{equation}\label{ext}
 n_{a}V^{a} = - n^{a}_{;\,a} = K_{uu}
 = n_{a}u^{a}_{;\,b}u^b =n_{a}f^a\equiv f\,,
\end{equation}
where $f^a=u^{a}_{;\,b}u^b=-fn^{a}$ is the acceleration vector of the
shell. Therefore, formula (\ref{eq-s1}) can be written as
\begin{equation}\label{eq-s1a}
  [\vec{n}\,r] = r[K_{uu}]=r [n_{a}f^a]\,.
\end{equation}

In order to obtain the equations of motion for the dust spherical
shell we shall  consider the normal variations of the hypersurface
$\Sigma$. Let each point $p \in\Sigma $ be displaced at a coordinate
distance $\delta x^{a}(p)=n^{a}\delta\lambda (p)$ in the direction of
the normal. As a result of the displacement, we obtain a new
hypersurface ${\tilde \Sigma}$. The initial and final positions of
the shell are fixed, therefore we have $\delta\lambda (p)=0,\ \forall
p\in\Sigma \cap\partial D ={\tilde\Sigma}_t \cap\partial D$. In
addition, we fix the metric $\gamma_{ab}$, and also all quantities on
$\Sigma$, so that $ \delta I_{m} = 0$.

As a result of displacement of the hypersurface $\Sigma $, the original
regions $D_{+}$ and $D_{-}$ are transformed into new regions ${\tilde
D}_{+} $ and ${\tilde D}_{-}$, such that ${\tilde D}_{-} \cup
{\tilde\Sigma} \cup {\tilde D}_{+} = D_{-} \cup \Sigma \cup D_{+} = D$.
Then, for example, the variation of the region $ D_{-} $ can be
represented as $\delta D_{-} = {\tilde D}_{-} \backslash D_{-} = D_{+}
\backslash {\tilde D}_{+}$. The change of the action (\ref{ig}) induced by
the displacement $\Sigma $, under the above conditions, is given by
\begin{equation}\label{deltaI}
   \delta I_{tot} = \delta I_{g} =
\int\limits_{{\tilde  D}_- \cup {\tilde D}_+}\!\! L_{g}\ d^2 x
  - \int\limits_{D_- \cup D_+}\!\! L_{g}\ d^2 x
 \cong - \int\limits_{\delta D_{-}}\left( L^{+}_{g}
  - L^{-}_{g}\right) d^2 x  \, .
\end{equation}
Here $ L^{+}_{g}$ and $ L^{-}_{g}$ are the Lagrangians determined by
the relation (\ref{Lg}) and calculated to the right and to the left
of $\Sigma$, respectively. Under the infinitesimal normal
displacement of the hypersurface $\Sigma$, the variation of the total
action takes the form
\begin{eqnarray}\label{deltaI2}
    \delta I_{tot} =
   - \int\limits_{\Sigma}\left( L^{+}_{g}
   - L^{-}_{g}\right) \delta x^{a} d \Sigma_{a} =
   \int\limits_{\Sigma} [L_{g}]\delta\lambda d\Sigma \, .
\end{eqnarray}
Hence, owing to the arbitrariness of $\delta\lambda $ and the
requirement $ \delta I_{tot}=0$, we find
\begin{equation}\label{[L]}
[L_{g}]= \left(L^{+}_{g} - L^{-}_{g}\right)|_{\Sigma}
=-\frac{c^3}{2\gamma}\,\sqrt{-\gamma}\ [2r(\vec{n} r)K_{uu}-(\vec{n}
r)^2]=0\, .
\end{equation}
Using formulas such as $[AB]=\bar A\,[B]+\bar B\, [A]$, where  $\bar
A = (A_{{|_{+}}} + A_{{|_{-}}})/2\, $, we obtain the equations of
motion in the form
\begin{equation}\label{eq-s3}
r\overline{(\vec{n} r)}\,[K_{uu}]+r[\vec{n} r] \overline{K_{uu}} -
\overline{(\vec{n} r)}\,[\vec{n} r]=0\,.
\end{equation}
After substitution of expressions for $[\vec{n} r]$ from
(\ref{eq-s1a}) into (\ref{eq-s3}), the equations of motion for the
dust spherical shell can be written as
\begin{equation}\label{midS}
\overline{K_{uu}}=\frac{1}{2}\,\left(
{K_{uu}}_{{|_{+}}}+{K_{uu}}_{{|_{-}}}\right) =
\overline{n_{a}f^a}=0\,.
\end{equation}

The relations (\ref{eq-s2}), (\ref{eq-s1a}) and (\ref{midS}) form the
necessary complete set of the boundary algebraic conditions imposed
on normal derivatives of the shell radius $(\vec{n}\,r)_{{|_{+}}} ,\
(\vec{n}\,r)_{{|_{-}}} $ and of the shell acceleration $f_{{|_{+}}}$
and $f_{{|_{-}}}$
(or component ${K_{uu}}_{{|_{+}}}$ and ${K_{uu}}_{{|_{-}}}$ of the
extrinsic curvature tensor on the shell) with respect to the internal
or external coordinates, respectively. In particular, this equations
imply
\begin{equation}\label{Snf}
    {K_{uu}}_{\pm} = n^{a}u_{a;\,b}u^{b}|_{\pm}
    = n^{a}\frac{Du_{a}}{ds}\,\biggl |_{\pm} = \mp \frac{k m}{2c^2 r^2}
\end{equation}
or
\begin{equation}\label{Snf0}
    {u^{a}_{;\,b}u^{b}}_{|_\pm} =
    \frac{Du^{a}}{ds}\,\biggl |_{\pm} = \pm\,\frac{k m}{2R^2}\, n^{a}
    \,.
\end{equation}
where $Du_{a}= u_{a;\,b}\,dx^{b}$ is the covariant differential.
These relations give us the equations of motions for the spherical
dust shell with respect to coordinates $x^a_{+}$ or $x^a_{-}$ of the
regions $D_{+}$ or $D_{-}$, respectively.

From the equations (\ref{Snf}) it follows  the two-dimensional
spherically symmetric analog of the the well-known Israel equations
\cite{israel1}
\begin{eqnarray}
      n^{a}{\frac{Du_{a}}{ds}}\biggl |_{+} & + &
      n^{a}{\frac{Du_{a}}{ds}}\biggl |_{-}  =  0\,,\label{4.39} \\
      n^{a}{\frac{Du_{a}}{ds}}\biggl |_{+} & - &
      n^{a}{\frac{Du_{a}}{ds}}\biggl |_{-} = - \frac{k m}{c^2 r^2}=
     - \frac{\chi \sigma}{2}\,.      \label{4.40}
\end{eqnarray}

\section{The reduced and effective actions for\\ the dust spherical shell}
\setcounter{equation}{0}

Now we can realize a reduction of system and construct the reduced
action for the shell. For this purpose we shall calculate action
$I_{tot} $ on the solutions of the vacuum Einstein equations
(\ref{gab})-(\ref{gik}) (or the equations
(\ref{eq-g1})-(\ref{eg-r})). In this case it is convenient to take
action $I_{tot}$ in the form (\ref{itot1}). In addition we shall take
into account the surface equations (\ref{eq-s2}), (\ref{eq-s1a}) and
(\ref{midS}). Note, that on this stage we explicitly use the
following consequences of these equations
\begin{eqnarray}\label{eqs}
      ~^{(4)}R =0\,, \qquad [\vec{n} r] =r[K_{uu}]\,, \qquad
      r^2 [K_{uu}]=-\frac{k m}{c^2}\,.
\end{eqnarray}
Substituting these relations into (\ref{itot1}) and taking into
account (\ref{idus2}), one finds
\begin{equation}\label{itot3}
{I_{tot}}_{| \{Eqs.~(\ref{eqs})\}} = J_{sh}+\acute{I_{\partial D}}+
I_{0}\,,
\end{equation}
where
\begin{eqnarray}\label{Lsh}
      && \qquad \qquad  J_{sh} = {I_{\Sigma}}_{| \{Eqs.~(\ref{eqs})\}}
      - mc\int\limits_{\gamma}\omega \nonumber \\
 &&  = -\biggl\{\int\limits_{\gamma}
  \biggl( mc+\frac{c^3}{2k}\, r^2 [K_{uu}]\biggl)\omega \biggl\}_{|\{Eqs.~(\ref{eqs})\}} =
        -\frac{1}{2}\int\limits_{\gamma}m c\, \omega
\end{eqnarray}
is the reduced action for the dust shell. This action must be
considered together with the boundary conditions (\ref{eq-s2}),
(\ref{eq-s1a}) and (\ref{midS}). The action $J_{sh} $ is quite
certain if in the neighborhood of $\Sigma$ the gravitational fields
are determined as the solutions of the vacuum Einstein equations
(\ref{gab}) - (\ref{gik}) which satisfy the boundary conditions
(\ref{eq-s2}), (\ref{eq-s1a}) and (\ref{midS}). The boundary term
$\acute{I_{\partial D}}={I_{\partial D}}_{| \{Eqs.~(\ref{eqs})\}} $
in (\ref{itot3}) now has a fixed value and not essentially for
further.

Note, that one usually comes to the action for the shell in the other
form. In our approach this form of the action can be found by partial
reduction of the initial action $I_{tot}$ when the last boundary
condition in (\ref{eqs}) is not taken into account. As a result, we
come to the action similar to the expression in braces in (\ref{Lsh})
or to some of its modification. Hence one can obtain the Lagrangian
of the shell in the frame of reference of the comoving observer.
However, the quantity $K_{uu} =n_{a}f^a = n_{a}u^{a}_{;\,b}u^b$
contains second derivatives of coordinates $x^a$ with respect to the
proper time of the shell. When these derivatives are eliminated by
integrating by parts, we obtain rather complicated Lagrangians and
Hamiltonians.

Now we introduce independent coordinates $x^{a}_{\pm}$ in each of the
regions $ D_{\pm}$. Then, the reduced action is the functional of
embedding functions $x^{a}_{\pm}(s)$ of the shell: $J_{sh} \equiv
J_{sh}[x^{a}_{-}(s), x^{a}_{+}(s)]$. Consider the variation of
integrand in $ J_{sh} $ with respect to these functions. We have
\begin{equation}\label{deltaw}
       \delta\omega_{|\gamma}^{\pm} = \delta\, ^{(2)}ds_{\pm} =
       \delta \left(\sqrt{\gamma_{ab}dx^{a}dx^{b}}\ \right)_{\pm}
      =- (u_{a;\,b}u^{b}\delta x^{a}\, ^{(2)}ds)_{\pm}
      + d(u_{a} \delta x^{a})_{\pm}\, .
\end{equation}
Hence, applying formulas $\delta^{a}_{b}=u^{a}u_{b}-n^{a}n_{b}$
and $u^a u_{a;\,b}=0$, we obtain
\begin{equation}\label{deltaw1}
       \left(\delta\, ^{(2)}ds - n^c u_{c;\,b}u^{b}n_a \delta x^{a}\,
       ^{(2)}ds\right)_{\pm} = d(u_{a} \delta x^{a})_{\pm}\, ,
\end{equation}
or, considering the boundary conditions (\ref{Snf}),
\begin{equation}\label{deltaw1a}
       \left(\delta\, ^{(2)}ds
       \pm\frac{k m}{2c^2 r^2}n_a \delta x^{a}\,
       ^{(2)}ds\right)_{\pm} = d(u_{a} \delta x^{a})_{\pm}\, .
\end{equation}
Further, using the conditions (\ref{n_u}) and (\ref{n_ua}), we have
\footnote{In the paper \cite{gl3} the factor $\sqrt{-\gamma}$ was
lost in these formulas. However, in the particular case of the
Schwarzschild solution it is of no importance since
$\sqrt{-\gamma}=1$}
\begin{equation}\label{un}
      (n_{a}\delta x^{a}ds)_{|\pm} =
      \{\sqrt{-\gamma}\ {(u^1 \delta x^0 -u^0 \delta x^1)ds}\}_{|\pm}
      = \{\sqrt{-\gamma}\ (dx^1 \delta x^0 -dx^0 \delta x^1)\}_{|\pm}\, .
\end{equation}
Therefore the variational formula (\ref{deltaw1}) takes the form
\begin{equation}\label{deltaw2}
      \left\{\delta\, ^{(2)}ds \pm\frac{k m}{2c^2 r^2}\sqrt{-\gamma}\
      (dx^1 \delta x^0 -dx^0 \delta x^1)\right\}_{\pm} =
      d(u_{a} \delta x^{a})_{\pm}\, .
\end{equation}

Now introduce the vector potential $B_{a}=B_{a}(x^0, x^1)$ with
the help of equation
\begin{equation}\label{poten}
      d\wedge (B_{a}dx^a ) \equiv G_{01} dx^0 \wedge dx^1 =
      -\frac{k m}{2c^2 r^2}\,\sqrt{-\gamma}\ dx^0 \wedge dx^1 \, ,
\end{equation}
where $G_{ab}\equiv B_{b,a}- B_{a,b}$. Note that, in
two-dimensional space, the integrability condition for the just
introduced relation holds identically. With this definition in
mind and owing to the fact that
\begin{equation}\label{deld}
  \delta (B_{a}dx^a ) -
  d(B_{a} \delta x^a)= G_{10}(dx^0 \delta x^1 -dx^1 \delta x^0) \, ,
\end{equation}
the variational formula (\ref{deltaw2}) can be written in the form
\begin{eqnarray}\label{deltaw3}
    \delta \left\{\, ^{(2)}ds \pm\ B_{a}dx^a\right\}_{\pm} =
      d\{(u_{a}\pm B_{a}) \delta x^{a}\}_{\pm}\, .
\end{eqnarray}
Thus, if we introduce the actions in the form
\begin{equation}\label{Ish}
      I^{\pm}_{sh} = - mc
   \int\limits_{\gamma} \left(\, ^{(2)}ds \mp B_{a}dx^a \right)_{|\pm}\, ,
\end{equation}
then, owing to the variational formula (\ref{deltaw3}), we shall
obtain the stationarity condition $\delta I^{\pm}_{sh}=0$ for the
fixed initial and final positions of the shell. The just obtained
actions are the natural modification of the action (\ref{Lsh}) which
is compatible with the boundary conditions. The stationarity
condition $\delta I^{\pm}_{sh}=0$ for arbitrary variation of
coordinates $x^{a}_{\pm}$ yield the equations of motion for the shell
with respect to external or internal coordinates. Therefore, formula
(\ref{Ish}) is the general form of the effective actions for the dust
spherical shell in general relativity, where the vector potential
$B_a $ is found from equation (\ref{poten}).

\section{The effective action for the spherical \\ dust shell}
\setcounter{equation}{0}

Now let us construct the effective actions for the spherical dust
shell in the Schwarzs\-child gravitational fields. Using curvature
coordinates, we choose common spatial spherical coordinates
$\{r,~\theta,~\alpha\}$ in $D_{\pm}$, and individual time coordinates
$t_{\pm}$ in $D_{\pm}$, respectively. Then the world sheet of the
shell $\Sigma$, in interior and exterior coordinates, is given by
equations $r=R_{-}(t_{-})$ and $r=R_{+}(t_{+})$, respectively.

The gravitational fields in the regions $D_{\pm}$ are described by
the metrics
\begin{equation}\label{Schwar}
      ~^{(4)}ds^{2}_{\pm}=f_{\pm}c^2 dt^{2}_{\pm}- f^{-1}_{\pm}dr^2
      -r^2 (d\theta^2+\sin^2 \theta d\alpha^2)\, ,
\end{equation}
where
\begin{equation}\label{f-sch}
      f_{\pm} = 1 - \frac{2k M_{\pm}}{c^2 r}\, ,
\end{equation}
and $M_{+}$ and $M_{-}$ are the Schwarzschild masses
$(M_{+}>M_{-})$.

In this case we have
\begin{equation}\label{pot-sc}
      B_{a}dx^a=c\varphi(t_{\pm},R)dt_{\pm}
      + U_{R}(t_{\pm},R)dR \, .
\end{equation}
Using the gauge condition $U_{R}(t_{\pm},R)=0 $,  the action
(\ref{Ish}) can be written as
\begin{equation}\label{4.2.4}
      I^{\pm}_{sh} =- mc \int\limits_{k}
      \left(^{(2)}ds \mp c\, \varphi dt\right)_{|\pm}\, .
\end{equation}
Further, formula (\ref{poten}) implies
\begin{equation}\label{dpot-sc}
      \frac{\partial \varphi}{\partial R}=\frac{k m}{2c^2 R^2}\,.
\end{equation}
From here, up to an additive constant, we find
\begin{equation}\label{pot-sc1}
      \varphi = - \frac{k m}{2c^2 R}\, .
\end{equation}
Eventually, the effective action for the shell can be represented
as
\begin{equation}\label{Ish1}
   I^{\pm}_{sh} = \int\limits_{\gamma}L^{\pm}_{sh}dt_{|\pm}
      =- \int\limits_{\gamma}
      \biggl(mc~^{(2)}ds \pm \frac{k m^2}{2 R} dt\biggr)_{|\pm}\, ,
\end{equation}
where
\begin{equation}\label{Lsh1}
      L^{\pm}_{sh} =
      -mc^2 \sqrt{f_{\pm}- f^{-1}_{\pm} R^2_{t\pm}/c^2}
      \pm U
\end{equation}
are the Lagrangians of the dust shell in the frames of reference
of stationary observers in the regions $D_{\pm}\, , \
(R_{t\pm}=dR/dt_{\pm})$, respectively, and
\begin{equation}\label{pot-en}
      U = - \frac{k m^2}{2R}
\end{equation}
is the effective potential energy of the shell gravitational
self-action.

It is easy to check that from the actions (\ref {Ish1}) the equations
of motion (\ref{Snf}) of the dust spherical shell follow.

The actions  $I^{\pm}_{sh}$ transform each into other under the
discrete gauge transformation $$
      M_{\pm} \stackrel{~}{\to} M_{\mp} \quad
      (f_{\pm}\stackrel{~}{\to}f_{\mp}), \quad
      U\stackrel{~}{\to} - \ U, \quad
      t_{\pm}\stackrel{~}{\to} t_{\mp}.
$$ They describe the transition from the interior observer to the
exterior one and vice versa.

Note that the actions $I^{\pm}_{sh} $ can be considered quite
independently. The regions $D_{\pm}$ together with the gravitational
fields (\ref{Schwar}) can also be regarded separately and
independently as manifolds with edges $\Sigma_{\pm}$. These edges
acquire the physical sense of the different faces of the dust shell
with the world sheet $\Sigma$ if the regions $D_{\pm}$ are joined
along these boundaries. The last can be realized only if the
conditions of isometry for the edges $\Sigma_{\pm}$
\begin{equation}\label{izo}
      f_{+}c^2 dt^{2}_{+}- f^{-1}_{+}dR^2 =
      f_{-}c^2 dt^{2}_{-}- f^{-1}_{-}dR^2 = c^2 d \tau^2 \, ,
\end{equation}
for the edges $\Sigma_{\pm}$ are fulfilled (or if the curves
$\gamma_{\pm}$ representing the world lines of the shell in the
coordinates $\{R,t_{\pm}\}$ coincide), where $\tau$ is the proper
time of the shell.

Consider some consequences following from the conditions of isometry for
the edges. First of all we have the relationships between the velocities
\begin{equation}\label{vel1}
     c^2 \frac{f_+}{R^2_{t+}} - \frac{1}{f_+} =
      c^2 \frac{f_-}{R^2_{t-}} - \frac{1}{f_-}\, ,
\end{equation}
\begin{equation}\label{vel2}
     R_{\tau}^2 \equiv \left(\frac{dR}{d\tau}\right)^2 =
     \frac{c^2 R^2_{t\pm}}{c^2 f_{\pm} - f^{-1}_{\pm} R^2_{t\pm}}\, , \qquad
     R_{t\pm}^2 \equiv \left(\frac{dR}{dt_{\pm}}\right)^2
     = \frac{c^2 f^{2}_{\pm} R^{2}_{\tau}}{c^2 f_{\pm} + R^2_{\tau}} \, .
\end{equation}

Further, from the Lagrangians $L^{\pm}_{sh} $ (\ref{Lsh1}) we find
the momenta and Hamiltonians for the shell
\begin{eqnarray}
      P_{\pm} = \frac{\partial L^{\pm}_{sh}}{\partial R_{t\pm}}
      = \frac{mR_{t\pm}}
      {f_{\pm} \sqrt{f_{\pm} - f^{-1}_{\pm} R^2_{t\pm}/c^2}}
      = \frac{m}{f_{\pm}} R_{\tau}\, , \label{imp}
\end{eqnarray}
\begin{eqnarray}
      H^{\pm}_{sh} = \frac{mc^2 f_{\pm}}
      {\sqrt{f_{\pm} - f^{-1}_{\pm} R^2_{t\pm}/c^2}}
      \mp U = mc^2 f_{\pm} \frac{dt_{\pm}}{d\tau}
      \mp U     \label{ham}
\end{eqnarray}
or
\begin{eqnarray}
      H^{\pm}_{sh} = c \sqrt{f_{\pm}(m^2 c^2 + f_{\pm}P_{\pm}^2)}
      \mp U  = mc^2 \sqrt{f_{\pm}
      + R^2_{\tau}/c^2} \mp U = E_{\pm}\, ,\label{ham1}
\end{eqnarray}
where $E_{\pm}$ are the shell energies which are conjugated to the
coordinate times $t_{\pm}$ and are conserved in the frames of reference of
the respective stationary observers (interior or exterior one).
Eliminating the velocity $R_{\tau}$ from (\ref{imp}) and (\ref{ham1}), the
conditions of isometry for edges can be written as
\begin{eqnarray}
      & f_{+}P_{+} = f_{-}P_{-} \, , \label{ipm-con}\\
      & \left(E_{+}+U\right)^2 -m^2 c^4 f_{+}
       = \left(E_{-}-U\right)^2 -m^2 c^4 f_{-}\, . \label{isom}
\end{eqnarray}
The last equation can be rewritten as
\begin{equation}\label{isom1}
  (E_{+}+E_{-})(E_{+}-E_{-}+2U)=m^2 (F_{+}-F_{-})\,.
\end{equation}
Substituting the expressions for $f_{\pm}$ and $U$ taken from
(\ref{f-sch}), (\ref{pot-en}) in this equation, we obtain
\begin{equation}\label{isom2}
  (E_{+}+E_{-})(E_{+}-E_{-})=\frac{k m^2}{R}\,(E_{+}+E_{-}-2(M_{+}+M_{-})\,)\,.
\end{equation}
Hence we find the relation between the Hamiltonians $H^{\pm}_{sh}$
and the Schwarzschild masses $M_{\pm}$
\begin{equation}\label{ham-con}
      H^{+}_{sh}= H^{-}_{sh} = (M_{+} - M_{-})c^2 = E \, .
\end{equation}
Here $E=E_{+}= E_{-}$ denotes the total energy of the shell, which is
conjugated both of the coordinate time $t_{+}$ and of that $t_{-}$
and whose value is independent of the stationary observer's position
(inside or outside of the shell). From now on  we shall treat the
relationships (\ref{ipm-con}) and (\ref{ham-con}), which appear in
the above independent description of the shell faces, as momentum and
Hamiltonian constraints. Thus, the dynamic systems with Lagrangians
$L^{\pm}_{sh}$ are not independent. They satisfy momentum and
Hamiltonian constraints (\ref{ipm-con}), (\ref{ham-con}) which ensure
of isometry of the shell faces.

The Lagrangians $L^{\pm}_{sh}$ (\ref{Lsh1}), as well as the relations
(\ref{vel1}) - (\ref{ham-con}), are valid only in a limited domain,
since the used curvature coordinates are valid outside the event
horizon only. Therefore, $L^{-}_{sh}$ can be used when $R>2k
M_{-}/c^2 $, and $L^{+}_{sh}$ for $R>2k M_{+}/c^2 \ \ (M_{+}>M_{-})$.

As is known, the complete description of the shells can be performed,
for example, in the Kruskal-Szekeres coordinates. With respect to
these coordinates the full Schwarzschild geometry consists of the
four regions $R^{+},\ T^{-},\ R^{-}, T^{+}$, detached by the event
horizons. Our above consideration concerned with the  $R^{+}$ region
only.

Formally, assuming $r$ to be the time coordinate, we can also use the
action for the shell in the form (\ref{Ish1}) under the horizon, i. e. in
the regions $T^{-}$- and $T^{+}$. In order to use the simplicity and
convenience of the curvature coordinates and to conserve the information
about the shells in the region $R^{-}$, it is sufficient to introduce an
auxiliary discrete variable $\varepsilon =\pm 1 $ and to make a change
$~^{(2)}ds_{\pm} \stackrel{~}{\to}\varepsilon_{\pm}~^{(2)}ds_{\pm} $ in
$I^{\pm}_{sh} $ (\ref{Ish1}) \cite{gl3}. Here, $\varepsilon_{\pm} =1 $
corresponds to the shell into the $R^{+}$-region, and $\varepsilon_{\pm}
=-1 $, to the shell into the $R^{-}$-region. Further, introduce the
quantities $\mu_{\pm}=\varepsilon_{\pm}m$. Then the extended action takes
the form
\begin{eqnarray}\label{Ish2}
   I^{\pm}_{sh}(\mu_{\pm}) =
      \frac{1}{2}\int\limits_{\gamma_{\pm}}L^{\pm}_{sh}(\mu_{\pm}) dt_{|\pm}
      =-\frac{1}{2}\int\limits_{\gamma_{\pm}}
      \left(\mu c~^{(2)}ds \mp U dt\right)_{|\pm}\, ,
\end{eqnarray}
where
\begin{equation}\label{Lsh2}
      L^{\pm}_{sh}(\mu_{\pm}) =
      - \mu_{\pm} c^2 \sqrt{f_{\pm}- f^{-1}_{\pm} R^2_{t\pm}/c^2}
      \pm U
\end{equation}
are the generalized Lagrangians describing the shell inside any of
the $R^{\pm}$-regions with respect to the curvature coordinates of
the interior $\{ t_{-}, R \}$ or exterior $\{ t_{+}, R \}$ regions.
The event horizons $R_g=2k M_{\pm}/c^2$, as before, are the singular
points of the dynamical systems (\ref{Ish2}) and must be excluded
from the consideration.

For the extended system (\ref{Ish2}) the Hamiltonians has the form
\begin{equation}\label{ham12}
      H^{\pm}_{sh}(\mu_{\pm}) =
      c\varepsilon_{\pm} \sqrt{f_{\pm}(m^2 c^2 + f_{\pm}P_{\pm}^2)}
      \mp U =
      \mu_{\pm}c^2 \sqrt{f_{\pm} + R^2_{\tau}/c^2} \mp U \, .
\end{equation}

Hence, taking into account the Hamiltonian constraints (\ref{ham-con}), we
find the standard relationships of the theory of the spherical dust shell
\cite{israel1} and rewrite them using new notation
\begin{eqnarray}
      \mu_{-} \sqrt{f_{-} + R^2_{\tau}/c^2} &-&
      \mu_{+} \sqrt{f_{+} + R^2_{\tau}/c^2} =
      \frac{k \mu^2}{Rc^2}\, ,   \label{4.2.22a}  \\
      \mu_{-} \sqrt{f_{-} + R^2_{\tau}/c^2} &+&
      \mu_{+} \sqrt{f_{+} + R^2_{\tau}/c^2} =
      2(M_{+}-M_{-})\, .  \label{4.2.22b}
\end{eqnarray}

Now consider briefly the self-gravitating shell when $ M_{-}=0$. Denote
$M_{+}= M $ and consider the shell moving in the $R^+ $-region. In the
exterior coordinates, the Lagrangian and Hamiltonian of the shell are of
the form
\begin{equation}\label{Lout}
      L^{+}_{sh} =
      -mc^2 \sqrt{1-\frac{2\gamma M}{c^2 R}-
      \left(1-\frac{2\gamma M}{c^2 R}\right)^{-1} \frac{R^2_{t+}}{c^2}}
      -  \frac{\gamma m^2}{2R} \, ,
\end{equation}
\begin{equation}\label{Hout}
      H^{+}_{sh} = c~\sqrt{1-\frac{2\gamma M}{c^2 R}}\
      \sqrt{m^2 c^2 + \biggl(1-\frac{2\gamma M}{c^2 R}\biggr)P_{+}^2}
      + \frac{\gamma m^2}{2R} \, .
\end{equation}
In the interior coordinates, the same shell is described by the
Lagrangian and Hamiltonian
\begin{equation}\label{Lin}
      L^{-}_{sh} = - mc^2 \sqrt{1- R^2_{t-}/c^2 } + \frac{\gamma m^2}{2R} \, ,
\end{equation}
\begin{equation}\label{Hin}
      H^{-}_{sh} = c\,\sqrt{m^2 c^2 + P_{-}^2}
                 - \frac{\gamma m^2}{2R} \, .
\end{equation}
The dynamical systems with $ L^{\pm}_{sh}$ obey the momentum and
Hamiltonian constraints $P_{-}= f_{+}P_{+}\,$,\ $ H^{+}_{sh} =
H^{-}_{sh} = Mc^2$, and are canonically equivalent in the extended
phase space \cite{gl3}. However they are not canonically equivalent
dynamic system, which is obtained at a choice of proper time as
evolutional parameter.

\section{Conclusions}
\setcounter{equation}{0}

In the paper, on the basis of the standard Einstein--Hilbert bulk
action and surface action for the dust the variational formalism for
a spherical dust shell is constructed. The total action also includes
the surface matching and boundary terms. The variational principle is
compatible with the boundary-value problem of the correspon\-ding
Euler--Lagrange equations. From the total action by variational
procedure the bulk equations and complete set of boundary conditions
are found. These equations are used for performing the reduction of
the system. As a result, we come to the reduced action (\ref{Lsh})
for the spherical dust shell which must be considered together with
surface equations. From reduced action we obtain the equations of
motion for the dust shell.

Further, by transforming the variational formula (\ref{deltaw}) for
the reduced action and  taking into account the surface equations
(\ref{Snf}), we obtain the effective action $I^{\pm}_{sh} $
(\ref{Ish}) for the dust shell which leads to correct equations of
motion. The above procedure is carried out in the curvature
coordinates for the interior and exterior regions $D_{\pm}$ of the
configuration. For the self-gravitating shell, the effective
Lagran\-gians $L^{\pm}_{sh}$ and Hamiltonians $H^{\pm}_{sh}$ describe
the gravitational collapse from the point of view of the interior
stationary observer and exterior remote stationary one.

The regions $ D_{\pm}\subset D$ together with the corresponding
gravita\-tional fields (\ref{Schwar}) can be treated as independent
submanifolds with ``loaded edges'' $\Sigma_{\pm}$ which can be
described by the actions $I^{\pm}_{sh} $. These edges acquire the
physical sense of different faces of the dust shell with the world
sheet $\Sigma$ if the regions $D_{\pm}$ are matched along these
boundaries. From the conditions of isometric equivalence of the edges
$\Sigma_{\pm}$ we obtain the momentum and Hamiltonian constraints
(\ref{ipm-con}), (\ref{ham-con}).

The effective Hamiltonian $H^{-}_{sh}$ was virtually postulated in
\cite{hajicek5} and was used for finding the energy spectrum of
quantum states of the dust shell with the bare mass $m$ that was less
than the Planck mass $m_{pl}$. In Ref. \cite{gl4} the Hamiltonians
$H^{\pm}_{sh}$ was used for constructing the quasi-classical model of
collapsing spherical configuration, for describing the tunneling
spherical dust shell, and also for the model of the pair creation and
annihilation of the shells. The method of constructing the effective
action is easily generalized to the case of more complex spherical
configurations with the space and surface distribution of fields and
matter. The present approach (see also \cite{gl3}) can be readily
generalized to the case of higher dimensions and can be used for
constructing the effective Lagrangians describing the cosmological
scenarios with branes. In that case, by using the variational
procedure, we can also find the complete set of boundary conditions
on the singular hypersurfaces, which are necessary  both in the
theory of brane worlds and in the shell theory (see, for example, the
papers \cite{muk} and \cite{dick} and references therein). In
conclusion, it should be stressed that, in contrast to \cite{gl3},
the approach taken in this paper is specially adapted for the
configurations which, after dimension reduction, are reduced to
2D-models. This allows us to use the equations specific to the
two-dimensional case only, simplifies the variational technique and
makes clearer the procedure of constructing the effective action.

\section*{Acknowledgements} I would like to express my gratitude to
M.P. Korkina for the fruitful discussion of the problems
considered in the paper.

\def\thesection{}
\renewcommand{\theequation}{A.\arabic{equation}}

\section{ Appendix A: The representation of the curvature scalar
in the two-dimensional space} \setcounter{equation}{0}

In the two-dimensional space there are specific relationships,
which can exist only in the spaces with dimensionality that equals
two. Some of them have been already written (see (\ref{2dim}) and
(\ref{ext})). Here we show that in the two-dimensional space the
curvature scalar is expressed (locally!) in terms of the
divergence of the vector constructed with the help of the vectors
of two-dimensional orthogonal basis $\{u^a , n^a \}$.

By definition we have
\begin{eqnarray}
  &u_{a\,;\,b\,;c} - u_{a\,;c\,;b}= R^{d}_{abc}u_d\,, \label{eq-u}\\
  &n_{a\,;\,b\,;c} - n_{a\,;c\,;b}= R^{d}_{abc}n_d \,.\label{eq-n}
\end{eqnarray}
Multiplying equation (\ref{eq-u}) by $u^f $, and equation
(\ref{eq-n}) by $n^f $ and applying formula
$u_{d}u^{f}-n_{d}n^{f}=\delta ^{f}_{d}$ gives
\begin{equation}\label{curva}
  R^{f}_{abc}=u^{f}(u_{a\,;b\,;c} - u_{a\,;c\,;b})
  - n^{f}(n_{a\,;b\,;c} - n_{a\,;c\,;b})\,.
\end{equation}
From here we find
\begin{equation}\label{rica}
  R_{ac}=u^{b}(u_{a\,;b\,;c} - u_{a\,;c\,;b})
  - n^{b}(n_{a\,;b\,;c} - n_{a\,;c\,;b})\,,
\end{equation}
\begin{equation}\label{scal}
  R=u^{b}(u^{a}_{\,;b\,;a} - u^{a}_{\,;a\,;b})
  - n^{b}(n^{a}_{\,;b\,;a} - n^{a}_{\,;a\,;b})
  = (u^{a}_{\,;b}u^{b} - u^{b}_{\,;\,b}u^{a}
  - n^{a}_{\,;b}n^{b} + n^{b}_{\,;b}n^{a})_{\,;a}\,,
\end{equation}
With the help of equations (\ref{2dim}) this formula can be
written in the form
\begin{equation}\label{scal1}
  R= 2(n^{b}_{\,;b}n^{a} - u^{b}_{\,;\,b}u^{a})_{\,;a} = 2V^{a}_{\,;\,a}\,,
\end{equation}
which gives the required equation (\ref{div}) expressing the
curvature scalar in terms of the divergence of vector $V^{a}$.


\end{document}